\documentclass[preprintnumbers,amsmath,amssymb,twocolumn,aps,prl]{revtex4-1}
\pdfoutput=1

\newcommand{\vev}[1]{\langle #1 \rangle}

\def\ga{\mathrel{\raise.3ex\hbox{$>$\kern-.75em\lower1ex\hbox{$\sim$}}}}
\def\la{\mathrel{\raise.3ex\hbox{$<$\kern-.75em\lower1ex\hbox{$\sim$}}}}

\newcommand\beq{\begin{equation}}
\newcommand\eeq{\end{equation}}
\newcommand\beqar{\begin{eqnarray}}
\newcommand\eeqar{\end{eqnarray}}
\usepackage{graphicx}
\usepackage{epsfig}
\usepackage{epsf}
\usepackage{rotating}
\usepackage{verbatim}

\usepackage{multirow}
\usepackage{natbib}
\usepackage{amsmath}
\usepackage{url}
\usepackage{hyperref}

\newcommand{\WMAP}{{\sc WMAP}}

\newcommand{\fnl}{$f_{\rm NL}$}

\begin{document}

\title{Non-Gaussianity Unleashed}

\author{Jonathan~S.~Horner and Carlo~R.~Contaldi}

\affiliation{Theoretical Physics, Blackett Laboratory, Imperial College, London, SW7 2BZ, UK}

\begin{abstract}
  The Hamilton-Jacobi (HJ) approach for exploring inflationary
  trajectories is employed in the generation of generalised
  inflationary non-Gaussian signals arising from single field
  inflation. Scale dependent solutions for \fnl\ are determined via
  the numerical integration of the three--point function in the curvature
  perturbation. This allows the full exploration of single field
  inflationary dynamics in the out--of--slow--roll regime and opens
  up the possibility of using future observations of non-Gaussianity
  to constraint the inflationary potential using model--independent
  methods. The distribution of `equilateral' \fnl\ arising from single
  field inflation with both canonical and non-canonical kinetic terms
  are show as an example of the application of this procedure.
\end{abstract}

\maketitle

Density perturbations due to primordial, Gaussian scalar metric
perturbations are a very good fit to the observed structure in the
Universe. The simplest model for seeding the initial, super--horizon
curvature fluctuations required in this picture is one where small
perturbations in a single inflaton field $\phi$ freeze out on large
scales during the phase of quasi--de Sitter expansion. The
perturbations are expected to be very close to Gaussian.

It is difficult to distinguish between different inflation models
if only scalar perturbations are measured. However if the non--Gaussian
contribution from higher order effects could be characterised it would
allow the breaking of the slow--roll degeneracy and resolve detailed
information about the shape of the inflaton potential $V(\phi)$ and in
principle allow the constraining of more complicated forms of the
action for $\phi$.

Over the past decade much work has gone into the quantitative
prediction of non--Gaussianity in both single field
\cite{Maldacena:2002vr}, multi--field models (see
\cite{2004PhR...402..103B} for a review) and ones with non--canonical
forms for the action . The formalism extends the calculation of the
two--point function in the gauge invariant curvature perturbation
$\zeta$ to third order. The two--point function, in the Fourier
expanded perturbation $\zeta({\mathbf k})$, defines the second moment
of the distribution of the perturbations via an isotropic power
spectrum $\langle\zeta({\mathbf k}_{1})\zeta^\star({\mathbf
  k}_{2})\rangle = (2\pi)^{3}\delta^{(3)}({\mathbf k}_{1} + {\mathbf
  k}_{2})P(k_{1})$. The three--point function vanishes in the Gaussian
case and is defined via an $f_{\rm NL}$ amplitude
\cite{Maldacena:2002vr}
\begin{eqnarray}\label{fnl_def}
\langle\zeta({\mathbf k}_{1})\zeta({\mathbf k}_{2})\zeta({\mathbf
  k}_{3})\rangle\! & = & \!\frac{6}{5}(2\pi)^{3}\delta^{(3)}({\mathbf k}_{1} +
{\mathbf k}_{2} + {\mathbf k}_{3})\times \\
&&\!\!\!f_{\rm
  NL}(k_1,k_2,k_3)
\left(P_{k_{1}}P_{k_{2}} + 2\,\text{perms}\right)\,.\nonumber
\end{eqnarray}
Analytical studies of the prediction for \fnl\ have focused on
particular configurations of the triangle ${\mathbf k}_{1} + {\mathbf
  k}_{2} + {\mathbf k}_{3}$ (`equilateral' $k_1=k_2=k_3$, `squeezed'
$k_1\ll k_2,\,k_3$, etc.) and on specific models of inflation. The
`squeezed' signal from single field inflation obeys a consistency
relation that means it is always small \fnl$\ll 1$, including in the non
slow--roll limit \cite{Creminelli:2004yq}. This level of
non--Gaussianity presents an observational challenge for even the most
optimistic future survey forecasts. Large levels of \fnl\ are instead
predicted in {\sl e.g.} the equilateral limits in single field models
in the out--of--slow--roll limit or with non-canonical kinetic terms
where the speed of sound $c_s$ for the inflaton can be less than unity
with respect to the speed of light
\cite{Tzirakis:2008qy,Noller:2011hd, Seery:2005gb}, DBI type models
\cite{Silverstein:2003hf}, and models with multiple fields that
undergo large accelerations \cite{Wands:2007bd}.

A number of authors have examined the numerical prediction in
out--of--slow--roll models that result in large \fnl\
\cite{Chen:2006xjb,Chen:2008wn}. These have focused on ad--hoc
modifications of the inflaton potential to induce temporary variations
in the slow--roll parameters $\epsilon$ and/or $\eta$ and have only
been applied to specific cases.  The subtleties involved in the
numerical evaluation of the integrals required in the {\sl in-in}
calculation of the three--point function (\ref{fnl_def}) were explored
in \cite{Chen:2006xjb,Chen:2008wn}.

This {\sl letter} introduces a generalised study of single field
non-Gaussianity employing the HJ trajectory formalism
\cite{PhysRevD.42.3936}. This procedure allows for the calculation of
\fnl\ arising from random trajectories and thus extends the HJ
approach for constraining the shape of the inflationary potential to
non-Gaussian observations. This type of analysis can complement the
similar exploration involving tensor contributions from different
inflationary trajectories. Two preliminary examples of these
explorations are shown here; the first being an ensemble of single
field inflation trajectories with a canonical kinetic term ($c_s=1$) and the second
being a case for $c_s\ne 1$. In both cases results for equilateral
\fnl\ as a function of wavenumber $k$ are presented. A more general
study and discussion of the method will be reported in
\cite{next}. A choice of units such that that $M^{2}_{\rm pl} \equiv
1/8\pi G = 1$ and $c=1$ is used throughout.

{\sl Hamilton-Jacobi formalism.---} The HJ formalism
\cite{PhysRevD.42.3936,Easther:2002rw} allows for the exploration of all possible 
inflationary trajectories consistent with a single inflaton 
evolving monotonically in an FRW universe. The natural basis for
labelling points in the phase-space of possible trajectories is an
infinite series of `slow--roll' co-ordinates that define the distance
of any point along the trajectory from pure de Sitter evolution. These
are defined as
\begin{equation}\label{SRdefn}
\epsilon  =  2\left(\frac{H'(\phi)}{H(\phi)}\right)^{2} \  \ \mbox{and} \ \ 
\lambda_l  =  2^{l} \frac{(H')^{l-1}}{H^{l}}\frac{d^{(l+1)}H}{d\phi^{(l+1)}}\,,
\end{equation}
where the value of the scalar field $\phi$ is used as an independent
parameter, $H$ is the Hubble rate, and primes denote differentiation
with respect to $\phi$. Re-casting the Friedmann
equations into a form where all quantities depend on $\phi$ one obtains 
\begin{eqnarray}\label{HamiltonJacobiEquations}
\frac{d\phi}{dt} & = & -2H'(\phi)\,,\\
\left[H'(\phi)\right]^{2} - \frac{3}{2}H(\phi)^{2} & = & -\frac{1}{2}V(\phi)\,,
\end{eqnarray}
where $t$ is cosmological time and $V(\phi)$ is the inflaton
potential. This allows one to define an infinite hierarchy of
differential equations whose solutions self--consistently determine possible
inflationary trajectories without the need to specify a potential
$V(\phi)$
\begin{eqnarray}
\frac{\mathrm{d}\epsilon}{\mathrm{d}N} & = & 2\epsilon(\epsilon -
\eta)\, ,\label{SR_diff_eqn1}\\
\frac{\mathrm{d}\lambda_l}{\mathrm{d}N} & = & (l\epsilon - (l-1)\eta)\lambda_l - \lambda_{l+1}\,\label{SR_diff_eqn2},
\end{eqnarray}
where $l=0,1,...$ and $\eta\equiv
\lambda_l$, $\xi\equiv \lambda_2$ and the
number of $e$-folds $N$, defined via $dN/dt=H$, has replaced $t$ as
the independent parameter by making use of the relation $d\phi/dN =
-\sqrt{2\epsilon}$. The system
(\ref{SR_diff_eqn1})-(\ref{SR_diff_eqn2}), together with the
differential equation $dH/dN=-H\epsilon$ can be integrated with
random initial conditions to obtain a Monte Carlo sampling of
inflationary trajectories \cite{Easther:2002rw}. The hierarchy can be
truncated consistently by imposing that $\lambda_l=0$ for all
$l>l_{\rm max}$. The resulting solutions are exact but
only cover a subset of all possible of solutions, limiting the
structure and complexity of the trajectories.

Briefly, the scheme imposed here for Monte Carlo sampling of HJ
trajectories is based on a `hierarchical prior' defined as follows;
the initial values of the parameters are $\epsilon = 1$, and the rest
of the slow--roll parameters are randomly drawn from a uniform
distribution with the ranges $[-0.1, 0.1]$ for $\eta$, and $[-0.1,
0.1]\times 10^{2-l}$ for $\lambda_l$, $l > 1$. The condition
$\lambda_{l_{\rm max}} = 0$ is imposed. The choice for $\epsilon$
ensures that inflation ends for that trajectory at the point where the
random boundary conditions are imposed. The system is then integrated
back in time for total of number of $e$-folds $N$ that is itself drawn
from a uniform distribution with range $[40, 80]$. This completely
fixes the inflationary model and, by construction, ensures inflation
both ends and provides the necessary $e$-foldings compatible with
observations. Other choices can be made for the priors used in drawing
random boundary conditions and also in the location of the conditions
along the trajectory and it should be noted that different choices will
affect the final distributions in the observables.

{\sl Two--point function.---} The trajectories define all background
quantities that are required for computing the correlations of $\zeta$
on super--horizon scales arising from the period of inflation. To do
this the late-time solution for $\zeta$ is obtained by integrating the
Mukhanov--Sasaki equation \cite{Mukhanov:1985rz,Sasaki:1986hm} arising
from expanding the action for $\zeta$ to second order in the
perturbations (with constant sound speed $c_{s}$):
\begin{equation}\label{MukhCode}
  \frac{\mathrm{d}^{2}\zeta_k}{\mathrm{d}N^{2}} + (3 + \epsilon - 2\eta)\frac{\mathrm{d}\zeta_k}{\mathrm{d}N} + \frac{c_{s}^{2}k^{2}}{a^{2}H^{2}}\zeta_k = 0\,,
\end{equation}
where $\zeta_k\equiv \zeta({\mathbf k})$ and $N$ is again used as the
independent variable. Bunch--Davies initial conditions
\cite{Bunch:1978yq} are assumed for each $\zeta_k$ at sufficiently
early times when $c_{s}k \gg a\,H$. The initial condition for $H$ is
chosen such that this condition holds at time $N = 0$. The system is
integrated numerically for a range of $k$ values up until the mode is
sufficiently larger than the horizon (i.e. $c_{s}k \ll a\,H$) to
ensure convergence of $\zeta_k$ to a constant.

The quantity $P_\zeta(k) = | \zeta_k|^2$, at a time when $c_{s}k \ll
a\,H$ is then the power spectrum of primordial, super--horizon
curvature perturbations that seeds structure formation. In the
slow-roll regime, where $\lambda_l \ll 1$, the dimensionless power
spectrum will approach a scale invariant solution $k^3\,P_\zeta(k)
\sim k^{n_s-1}$. The $n_{s}$ for each trajectory is calculated using a
second order finite difference scheme around a pivot point
$k_\star$. The first order slow--roll prediction is $n_s-1\approx
2\eta-4\epsilon$ \cite{Stewart:1993bc}. $P_\zeta(k)$ can then be
compared to observations to constrain the possible set of solutions
$\lambda_l(N)$ or equivalently $H(N)$ and/or $V(\phi)$. The latter
being the fundamental property of interest in determining the nature
of the inflaton.

{\sl Three-point function.---} The inflationary trajectories can
also be used to obtain generalised non--Gaussian predictions. The
amplitude of the three--point correlation function for $\zeta$ can be
obtained by considering the action for a perturbed, minimally coupled
scalar field up to third order in the perturbation\cite{Maldacena:2002vr, Noller:2011hd, Seery:2005gb}
\begin{eqnarray}\label{action}
  S &=& \!\!\int d^4x\, \frac{a^{3}\epsilon}{c_{s}^{2}}\left[-\frac{2}{3H}\left(1 - \frac{1}{c_{s}^{2}}\right)\dot{\zeta}^{3}\right.\nonumber\\
 & & \!\!\!\!\left. + \frac{1}{c_{s}^{2}}\left(3(c_{s}^{2} - 1) + \epsilon + 2\eta\right)\zeta \dot{\zeta}^{2} + \frac{1}{a^{2}}(1 - c_{s}^{2} + \epsilon)\zeta(\partial\zeta)^{2}\right.\nonumber\\
 & &\!\!\!\!\left.  - \frac{\epsilon}{a^{2}}(\epsilon - \eta)\zeta^{2}\partial^{2}\zeta - \frac{2\epsilon}{c_{s}^{2}}\left(1 - \frac{\epsilon}{4}\dot{\zeta}\partial_{i}\zeta\partial_{i}\zeta\partial_{i}\partial^{-2}\dot{\zeta}\right)\right.\nonumber\\
 & & \!\!\!\!\left. + \frac{\epsilon^{2}}{4c_{s}^{2}}\partial^{2}\zeta\partial_{i}\partial^{-2}\dot{\zeta}\partial_{i}\partial^{-2}\dot{\zeta}\right]\,,
\end{eqnarray}
where overdots denote differentiation with respect to $t$,
$\partial_i\equiv \partial/\partial x_i$, and $\partial^2$ and
$\partial^{-2}$ are the Laplacian operator and its inverse. 
This action differs from ones usually presented in the literature in
that no field redefinition has been introduced in order to avoid
assumptions about boundary conditions that would lead to unphysical
evolution of some of the terms for generalised trajectories i.e. ones
where $\eta_{V} = 2(\epsilon - \eta)$ is still evolving when $\zeta_k$ is outside the
horizon. This will be elaborated on in \cite{next}. The three--point function at the end of inflation with $N=N_e$ can
be computed in the interaction picture by integrating the interaction
Hamiltonian in time \cite{Maldacena:2002vr,Seery:2005gb}
\begin{eqnarray}\label{interaction}
  &&\!\!\!\!\vev{\zeta_{k_1}(N_e)\zeta_{k_2}(N_e)\zeta_{k_3}(N_e)} =\nonumber\\
   &&\!\!\!\!-i \int^{N_e}_{-\infty}\!\!\!\! \, dN \vev{\left[\zeta_{k_1}(N_e)\zeta_{k_2}(N_e)\zeta_{k_3}(N_e),H_{\rm int}(N)\right]}\,,
\end{eqnarray}
where $H_{\rm int}$ is the Hamiltonian for the three--point function arising from (\ref{action}).

This calculation can be carried out analytically in the slow-roll
regime \cite{Seery:2005gb, Maldacena:2002vr} and in the out of slow--roll picture
and/or non-canonical field case if the slow-roll parameters are
constant \cite{Noller:2011hd,Ribeiro:2012ar}. When considering the out of
slow--roll case with general, non--stationary trajectories the
calculation must necessarily be carried out numerically. A numerical
calculation of the integrals involved in (\ref{interaction}) has been
carried out for specific inflation models with features in the
potential \cite{Chen:2006xjb,Chen:2008wn}. Here (\ref{interaction}) is
integrated numerically for each Monte Carlo trajectory \footnote{ A
  detailed review of the numerical integration and the subtleties it
  involves will be given in \cite{next}.} The equilateral case of
$k_{1} = k_{2} = k_{3} \equiv k$ is chosen in what follows but the
calculation is valid for all configurations. The aim in this case is
to obtain the function \fnl$(k)$.

\begin{figure}[t]
  \begin{center}
    \includegraphics[width=9cm,trim=0cm 5cm 0cm 3cm,clip]{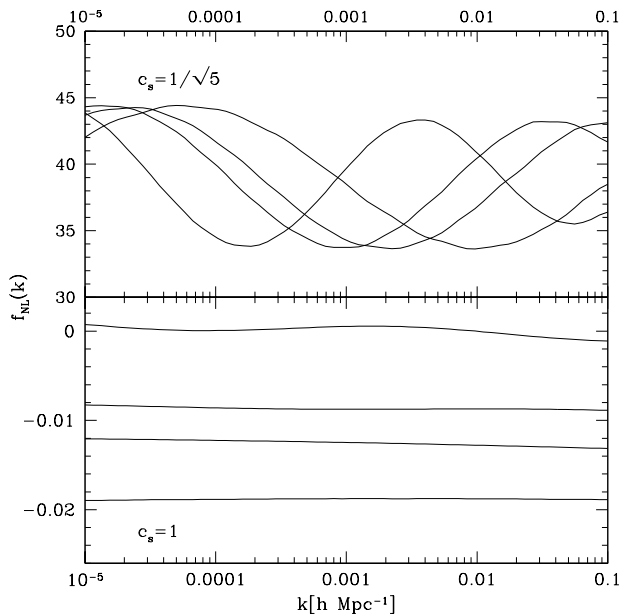} 
    \caption{A selection of \fnl$(k)$ spectra arising from typical
      trajectories for two cases; constant $c_s=1$
      (bottom) and $c_s=1/\sqrt{5}$ (top).  The \fnl\ in the $c_s\neq
      1$ exhibits strong scale dependence where as for $c_{s} = 1$ it is more mild.}
    \label{fig:fnl}
  \end{center}
\end{figure}

From equations (\ref{fnl_def}) and (\ref{interaction}), this can be
re-cast as \fnl$(k) =
2\,\text{Im}(z_k)$ where the complex mode $z_k$ satisfies the following
differential equation
\begin{eqnarray}\label{Z_ode}
\frac{\mathrm{d}z_k}{\mathrm{d}N} &=&
\left(\frac{1}{\zeta^{\star}_{k}}\frac{\mathrm{d}\zeta_{k}^{\star}}{\mathrm{d}N} -
  \frac{2}{\zeta_{k}}\frac{\mathrm{d}\zeta_{k}}{\mathrm{d}N}\right)z_k +
f_{1}\frac{\zeta_{k}^{\star}}{\zeta_{k}^{2}}\left(\frac{\mathrm{d}\zeta_{k}}{\mathrm{d}N}\right)^{3}
+\nonumber\\
&&f_{2}|\zeta_{k}|^{2} + f_{3}\frac{\zeta_{k}^{\star}}{\zeta_{k}}\left(\frac{\mathrm{d}\zeta_{k}}{\mathrm{d}N}\right)^{2}\,,
\end{eqnarray}
with initial condition $z_k\rightarrow 0$ as $N \rightarrow -\infty$
and $f_{i}$ are defined as
\begin{eqnarray}
f_{1} & = & \frac{10Ha^{3}\epsilon}{9c_{s}^{2}}\left(1 - \frac{1}{c_{s}^{2}}\right)\,,\nonumber\\
f_{2} & = & -\frac{5k^{2}a\epsilon}{6Hc_{s}^{2}}\left(c_{s}^{2} - 1 + \epsilon - 2\eta\right)\,,\\
f_{3} & = & -\frac{5Ha^{3}\epsilon}{6c_{s}^{4}}\left(6(c_{s}^{2} - 1) + 4\eta + \frac{3\epsilon^{2}}{4}\right)\,.\nonumber
\end{eqnarray}
The numerical integration of the system requires some care due to the
highly oscillatory nature of the integrand at early times and the
evolution of slow--roll parameters throughout the
integration. Convergence of the integrals with respect to both early
and late time contributions has been verified \cite{next}.

Two examples of the results obtained from the numerical procedure
discussed above are reported here. The first is an ensemble of ${\cal
  O}(10^5)$ trajectories with canonical speed of sound $c_s=1$. The
second is for a case where $c_s\ne 1$. For simplicity a constant value
$c_s=1/\sqrt{5}$ is chosen. From a Lagrangian perspective this
corresponds to changing the coefficient and power of the kinetic term
to functions of $c_{s}^{2}$. To include $c_{s}$ evolution an extra set of slow--roll parameters needs to be considered and consequently the functions $f_{i}$ are more complicated. 

\begin{figure}[t]
  \begin{center}
    \includegraphics[width=9cm,trim=0cm 5cm 0cm 3cm,clip]{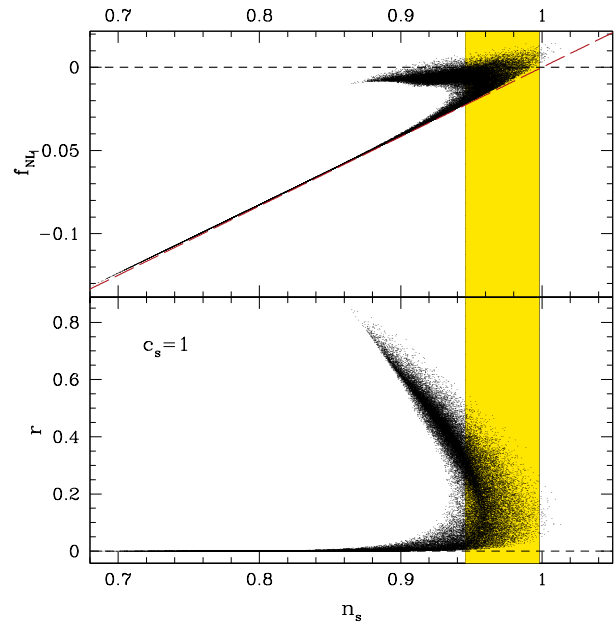} 
    \caption{The values of the tensor-to-scalar ratio $r$ (bottom) and
      \fnl\ (top) plotted against the scalar spectral index $n_s$. All
      quantities are obtained from the numerical spectra at a
      single pivot scale. Some $10^{4}$ trajectories were included in
      the ensemble with $c_s=1$. The shaded area shows the 95\%
      confidence region for $n_s$ allowed by the \WMAP\ 9-year results
      \cite{Hinshaw:2012fq}. The red (dashed) line shows \fnl $\sim
      5(n_{s} - 1)/12$ attractor expected in the slow--roll
      regime for equilateral \fnl.}
    \label{fig:nsr}
  \end{center}
\end{figure}

\begin{figure}[t]
  \begin{center}
    \includegraphics[width=9cm,trim=0cm 5cm 0cm 3cm,clip]{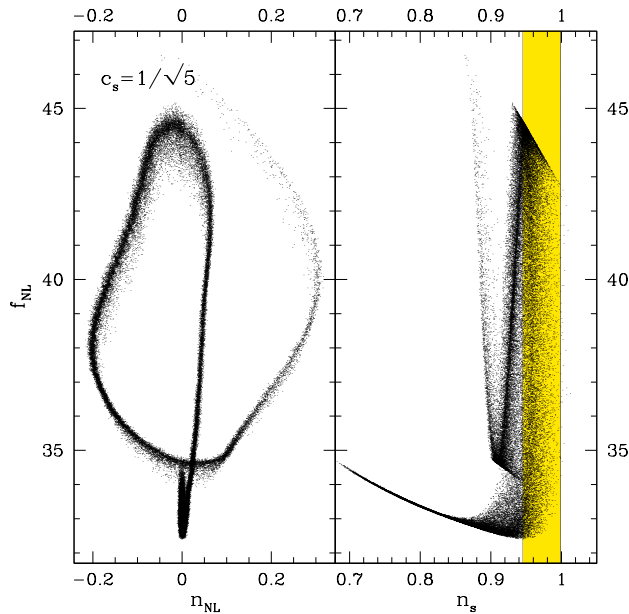} 
    \caption{\fnl\ plotted against $n_s$ (right) and $n_{\rm
        NL}$ (left) for the same `slow--roll' trajectories as in
      Fig.~\ref{fig:nsr} but with constant $c_s=1/\sqrt{5}$. The
      amplitude of the non-Gaussianity is much larger than for the
      $c_s=1$ case. The complicated shape of the attractor is a
      consequence of the sinusoidal nature of $f_{\rm NL}(k)$ which is
      not well approximated by a power law.}
    \label{fig:cs5}
  \end{center}
\end{figure}

The integration is carried out with the aid of GPU parallelisation and
only requires on the order of a few minutes for the \fnl\ at all $k$
samples to be computed for the entire ensemble. \fnl\ spectra
resulting from typical trajectories with $n_s\approx 1$ for both
ensembles are shown in Fig.~\ref{fig:fnl}. The amplitude of \fnl\ for
the canonical case is small, in agreement with analytical estimates
for single field inflation \cite{Maldacena:2002vr} and mild scale
dependence is present in some of the solutions. The \fnl\ for the
non--canonical case have a much higher amplitude, in agreement with
analytical calculations \cite{Noller:2011hd} for the constant
out--of--slow--roll case, and display a sinusoidal scale dependence in
$\ln k$.

The phase portrait of the canonical ensemble is shown in
Fig.~\ref{fig:nsr} where the tensor--to--scalar ratio $r$ and \fnl\
values are plotted against the $n_s$ for each trajectory. The values
shown are all obtained from the numerical spectra and are sampled at a
fixed pivot scale $k_\star$ chosen to be to the largest mode in our
sample corresponding to the scale of the horizon today.

The distribution of $r$ is as expected with a clear slow--roll
`attractor', this is due to the fact that for the $k_\star$
corresponding to large scales today the spectra are sampled early on
in the inflationary trajectories where the slow--roll parameters are
typically small given our choice of hierarchical priors at the {\sl
  end} of inflation. A more complex picture may arise from a wider
prior, increasing $l_{\text{max}}$ or for a choice of boundary
conditions at the {\sl start} of inflation. The results also show a
clear attractor in the value of \fnl$(k_\star)$ that agrees with the
slow--roll limit relation \fnl$\sim \frac{5}{12}(n_{s}-1)$ for the
equilateral case \cite{Maldacena:2002vr}.

The phase picture for the non--canonical case is shown in
Fig.~\ref{fig:cs5}. The amplitude of \fnl\ at the pivot point is
plotted against $n_s$ and the spectral tilt of \fnl\ defined as $n_{\rm
  NL} = \mathrm{d}\ln |f_{\rm NL}|/\mathrm{d}\ln k$ at the pivot
scale. The attractor in this case displays a complicated
structure. This is due to the fact that, for trajectories that result
in near scale invariant scalar power spectra, \fnl\ is strongly scale
dependent and is not well described by a power law amplitude and index
\footnote{The fact that the \fnl\ {\sl vs} $n_{\rm NL}$ portrait
  resembles the greek letter $\phi$ appears to be entirely
  coincidental.}. A better parametrisation of \fnl\ in this case may
be of the form \fnl$(k)= f_{\rm NL}^0 (1+g\,\sin(\omega
\ln k))$. Observations from {\sl Planck} and other future surveys should
be able to constrain the amplitudes of \fnl\ seen in this case
\cite{Burigana:2010hg} and potentially detect any scale dependence of
the type seen in these solutions.

These results show that it is possible to compute \fnl\ for thousands
of single field inflation models along with the usual $n_{s}$ and
$r$. The method agrees with the expected results of \fnl\ for
canonical single field slow--roll inflation, i.e. small and probably
unobservable. However, allowing non--canonical $c_s$ generally results
in observable levels of \fnl\ with significant scale
dependence. Allowing for more structure and widening priors used in
sampling the trajectories may also result in an extended range of
amplitudes for \fnl. The method will be useful for constraining
general single field inflation potentials and/or Lagrangian densities
(see
e.g. \cite{2003ApJS..148..213P,2008JCAP...01..010L,Bean:2008ga}). Further
work will explore the landscape of models and, in particular, extend
the approach to different `shapes' or configurations for $k_{1}$,
$k_{2}$, and $k_{3}$ and make detailed comparisons to observations.

Johannes Noller is acknowledged for useful discussions. This research is
supported by an STFC student grant ST/F007027/1 and Consolidated grant
ST/J000353/1.

\bibliography{paper}

\end{document}